\documentclass[12pt]{article}

\usepackage{amsmath}
\usepackage{color}
\usepackage{epsfig,graphics}
\usepackage{amssymb}
\usepackage{latexsym}
\usepackage[super,sort&compress]{natbib}

\newcommand{\MW}{\mathcal{W}}
\newcommand{\MH}{\mathcal{H}}
\newcommand{\MG}{\mathcal{G}}

\newcommand{\la}{\langle}
\newcommand{\ra}{\rangle}

\newcommand{\sech}{\textrm{sech}}
\newcommand{\tO}{\textrm{otherwise}}

\newcommand{\hC}{\hat{\chi}}
\newcommand{\Tr}{\textrm{Tr}}

\def \be {\begin{equation}}
\def \ee {\end{equation}}
\def \bea {\begin{eqnarray}}  
\def \eea {\end{eqnarray}}

\title{Bounds on integrals of the Wigner function: the hyperbolic case}
\author{J. G. Wood and A. J. Bracken \\
Centre for Mathematical Physics, University of Queensland, \\
Brisbane, Australia 4072}

\begin{document}

\maketitle

\begin{section}{Abstract}

Wigner functions play a central role in the phase space formulation of quantum
mechanics. Although closely related to classical Liouville densities, Wigner
functions are not positive definite and may take negative values on subregions
of phase space. We investigate the accumulation of these negative values by
studying bounds on the integral of an arbitrary Wigner function over noncompact
subregions of the phase plane with hyperbolic boundaries. We show using
symmetry techniques that this problem reduces to computing the bounds on the
spectrum associated with an exactly-solvable eigenvalue problem and that the
bounds differ from those on classical Liouville distributions. In particular,
we show that the total ``quasiprobability'' on such a region can be greater
than 1 or less than zero. 

\end{section}

\begin{section}{INTRODUCTION}

Since its introduction \cite{Wigner32},  the Wigner function has been the
subject of extensive study in the fields of quantum physics, quantum chemistry
and signal analysis (see \cite{Groenewold46,Moyal49, Hillery84,
Schleich01,Mori62, Carruthers83,Cohen89, Cohen95,Williams96} and references
therein). Since Wigner functions represent quantum states on phase space, they
play a key role in the phase space formulation of quantum mechanics. They are
also designed to closely resemble the joint densities of position and momentum,
known as Liouville densities, that are used in classical mechanics. In quantum
physics, such studies have been stimulated in recent times by the development
of quantum tomography, which has enabled the reconstruction of  Wigner
functions corresponding to states of a variety of quantum systems
\cite{Raymer97}. Such experimental observations have confirmed that Wigner
functions can be negative on subregions of phase space.  This is one of several
properties that can be used to distinguish Wigner functions from classical
Liouville densities.

The study of these ``quantum properties'' has been approached in a number of
ways including calculations of pointwise bounds on Wigner functions and bounds
on various moments \cite{Baker58,Price65,DeBruijn67,Janssen81,Lieb90}. A more
recent development has been the study of bounds on integrals of the Wigner
function over subregions of the phase space \cite{Bracken99, Bracken02,
Bracken03A}, which we denote by $\Gamma$ . We call such integrals {\em
quasiprobability integrals} (qpis). For a given subregion $S$ of $\Gamma$, the
problem of determining best possible upper and lower  bounds on all possible 
qpi's over $S$  has been shown to be equivalent to the problem of determining
the supremum and infimum of the spectrum of the \emph{region operator}
associated with $S$. This operator is just the image under Weyl's quantization
map \cite{Weyl31} of the characteristic function of $S$, namely the function
that equals $1$ on $S$ and $0$ elsewhere on $\Gamma$. In the special case of a
quantum system with one linear degree of freedom, it has been shown that for
any subregion of the phase plane enclosed by an ellipse, the eigenvalue problem
is exactly solvable and the bounds on qpi's can be obtained analytically for
ellipses of arbitrary size \cite{Bracken99}.

The determination of bounds on qpis is important not only because it provides
information about the structure of theoretically  possible Wigner functions,
which is a question of mathematical interest, but also because an understanding
of that structure provides checks on experimentally determined Wigner
functions.  It is therefore  of interest to know if there are other subregions
of the phase plane, and more generally of phase space, for which the spectrum
of the associated region operators, and hence the best possible upper and lower
bounds on all possible associated qpis, can be determined exactly. In this
paper, we show that an exact formula for the spectrum of the region operator,
from which the bounds are easily obtained numerically, can be derived for
subregions of the phase plane with hyperbolic symmetry. The solvability of the
eigenvalue problems for the corresponding region operators, as in the case of
elliptical subregions discussed earlier, relies on the invariance of these
regions under one-parameter subgroups of the metaplectic group
$Mp(2,\mathbb{R})$ of transformations of the phase plane.  This group consists
of all real transformations of the form
\begin{equation}
T : (q,p) \rightarrow (q',p') \,=\, (\alpha q + \beta p + q_0, \gamma q + \delta
p + p_0)\,, 
\label{metagroup}
\end{equation}
where $\alpha \delta - \beta \gamma=1$. In this paper, the subgroup of
$Mp(2,\mathbb{R})$ formed by the transformations $T_\sigma: (q,p) \rightarrow
(\sigma q, p/\sigma)\,, \sigma >0$ is of particular importance.

Several illustrative examples of eigenvalue problems for hyperbolic regions
are considered in what follows, including the interesting limiting case of  an
infinite wedge. We shall be concerned with  quantum systems with one linear
degree of freedom, described in terms of  a Hilbert space of states $\MH$, and
with  the properties of Wigner functions on the associated $(q,p)$ phase plane
$\Gamma$. We are not concerned with dynamics, and consider each Wigner function
at a fixed time. Dimensionless phase plane coordinates $(q,p)$ are used, and
in effect we set $\hbar=1$. Finally, we note that in the absence of limits of
integration, integrals are assumed to run from $-\infty$ to $\infty$.

\end{section}

\begin{section}{BOUNDS ON QUASIPROBABILITY INTEGRALS}

The Wigner function corresponding to a pure state $\psi \in \MH$ has the
definition
\begin{equation}
W_\psi(q,p) \,=\, \frac{1}{\pi} \int \overline{\psi}(q+\tau)\psi(q-\tau) e^{2ip
\tau} d\tau.
\label{bqi01}
\end{equation}
For a mixed state, the Wigner function is a convex linear combination of such
integrals.   It is known that Wigner functions are bounded at every point
$(q,p) \in \Gamma$ such that $-1/\pi \leq W(q,p) \leq 1/\pi$ and that they
satisfy the normalization conditions
\begin{equation*}
\int_\Gamma W dq dp \,=\,1 \,, \quad 0 \,\leq\, \int_\Gamma W^2 dq dp
\,\leq\, \frac{1}{2\pi}\,,
\end{equation*}
where the value $1/{2\pi}$ is attained if and only if $W$ corresponds to a pure
state.

More generally, an operator $\hat{A}$ is unitarily related to a phase space
function $A(q,p)$ by the Weyl-Wigner transform \cite{Dubin00} and
its inverse,
\begin{equation}
A={\cal W}({\hat A})\,,\quad {\hat A}={\cal W}^{-1}(A)\,.
\label{WWtransform}
\end{equation}
Here ${\cal W}^{-1}$ is Weyl's quantization map and $\mathcal{W}$ is such that
the Wigner function corresponding to a quantum density operator $\hat{\rho}$ is
given by $W_\rho= {\cal W}({\hat \rho})/{(2\pi)}$.  In this paper we make
extensive use of the configuration realization, in which $\hat{A}$ can be
expressed as an integral operator
\begin{equation}
(\hat{A} \psi)(x) \,=\, \int A_K(x,y) \psi(y) dy\,.
\label{bqi03}
\end{equation}
We refer to the function $A_K(x,y)$ as the configuration kernel of $\hat{A}$.
It is related to the phase space function $A(q,p)$ by the formulas
\cite{Osborn95,Bracken03B}
\begin{eqnarray}
A(q,p) &=& \int A_K(q-y/2,q+y/2) e^{ipy}dy\,, \quad \label{bqi04a} \\ 
A_K(x,y) &=& \frac{1}{2\pi}\int A((x+y)/2,p) e^{ip(x-y)} dp \,, \label{bqi04b}
\end{eqnarray}
which provide an explicit realization of the transformations (\ref{WWtransform}).  

An important property of Wigner functions is that quantum averages on phase
space take the same form as classical averages: if $A(q,p)$ is the phase space
representation of a quantum observable ${\hat A}$,  then its quantum average in
the state with density operator ${\hat \rho}$ and corresponding Wigner function
$W_\rho$ is given by
\begin{equation}
\la {\hat A} \ra \,= \Tr(\hat{A}\hat{\rho})\,=\, \, \int_\Gamma W_\rho(q,p)
A(q,p) dq dp \,.
\label{bqi05}
\end{equation}

The qpi of a Wigner function $W$ over a subregion $S$ of $\Gamma$ may be written
as the functional
\begin{equation}
Q_S[W] \,=\, \int_S W(q,p) dq dp \,.
\label{bqi06}
\end{equation}
Note that the integral on the RHS can be rewritten in terms of the
characteristic function  $\chi_S(q,p)$ that equals $1$ on $S$ and $0$ on its
complement: 
\begin{equation*}
Q_S[W] \,=\, \int_\Gamma  W(q,p)  \chi_S (q,p) dq dp\,,
\end{equation*}
and, by comparing with (\ref{bqi05}), we can write
\begin{equation}
Q_S[W] =\, \la \hC_s \ra\,,
\label{characave}
\end{equation}
where we have introduced the \emph{region operator} $\hC_S = \MW^{-1}(\chi_S)$
\cite{Bracken99}, with configuration kernel (as given by (\ref{bqi04b}))
\begin{equation}
\chi_{S,K}(x,y) = \frac{1}{2\pi}\int \chi_S((x+y)/2,p) e^{ip(x-y)} dp\,.
\label{bqi08}
\end{equation}

Since the expectation value of a quantum operator always lies between the
extremal values of its spectrum, we deduce from (\ref{characave}) that $Q_S[W]$
must lie between the infimum and the supremum of the spectrum of $\hC_S$.
Moreover, as the spectral bounds on the expectation value of an operator can
be approached arbitrarily closely with normalized states in $\MH$, these bounds
are best-possible. Hence the best-possible bounds on the qpi functional $Q_S$
are provided by the extremal solutions to the integral equation
\begin{equation}
(\hC_S \psi)(x) \,=\, \int \chi_{S,K}(x,y) \psi(y) dy \,=\, \lambda \psi(x)\,
\label{bqi09}
\end{equation}
that defines the eigenvalue problem for $\hC_S$.

For a general region $S$, the integral equation (\ref{bqi09}) is not exactly
solvable and the bounds on its spectrum must be obtained by using computational
methods. However, there is a subclass of regions for which the (generalized)
eigenvalues and eigenfunctions can be determined exactly. This subclass is the
set of regions that are each invariant under a one-parameter subgroup of the
metaplectic (or linear canonical) group of transformations (\ref{metagroup}) of
the phase plane. Any such transformation $U$  has the special property that its
inverse Weyl-Wigner transform ${\hat U}= {\cal W}^{-1}(U)$ is a unitary (and
thus spectrum preserving) operator acting on $\MH$. If a subregion $S$ of
$\Gamma$ is invariant under a metaplectic transformation, then it follows that
the associated region operator $\hC_S$ is invariant under the corresponding
unitary operation $\hat{U}$, that is generated by an operator $\hat{r}$ of no
greater than  the second degree in $\hat{q}$ and $\hat{p}$. It follows  that
$[\hC_S,\hat{r}]=0$ and hence that the eigenfunctions of $\hC_S$ may be chosen
so that they are also eigenfunctions of $\hat{r}$. These are readily obtained
by solving the eigenvalue problem for $\hat{r}$.

This approach can be applied to regions that are bounded by ellipses,
hyperbolas, parabolas and straight lines. (If  the boundary is composed of
several curves, then each curve must be invariant under the same one-parameter
subgroup of $Mp(2,\mathbb{R})$.)  In the case of elliptical regions, the best
possible bounds have already been  described \cite{Bracken99}, while the fact
that the marginal distributions of the Wigner function are true probability
density functions \cite{Bertrand87} implies that integrals over regions bounded
by parallel straight lines must lie in the interval $[0,1]$. In this paper, we
consider the problem of determining the best possible bounds on qpis over
regions with hyperbolic boundaries.

\end{section}

\begin{section}{BEST POSSIBLE BOUNDS ON QPIS FOR HYPERBOLIC REGIONS}

In order to demonstrate our technique for constructing the bounds on qpis, we
begin with a simple example. Let $C_k$ be the hyperbolic curve consisting of
all points that satisfy
\begin{equation}
qp \,=\, k\,,\quad k\geq 0\,,
\label{bpb01}
\end{equation}
as depicted on in part (a) of Figure \ref{onecurve}. Note that $C_k$ is itself
composed of two curves, namely $C_k^+$, which lies in the positive $(q,p)$
quadrant of $\Gamma$ and $C_k^-$, which lies in the negative $(q,p)$ quadrant
of $\Gamma$. It is clear that the curves $C_k^\pm$ are separately
invariant under the action of the transformation $T_\sigma: (q,p) \rightarrow
(\sigma q, p/\sigma)$ for all $\sigma>0$.

\begin{figure}
\centerline{\psfig{figure=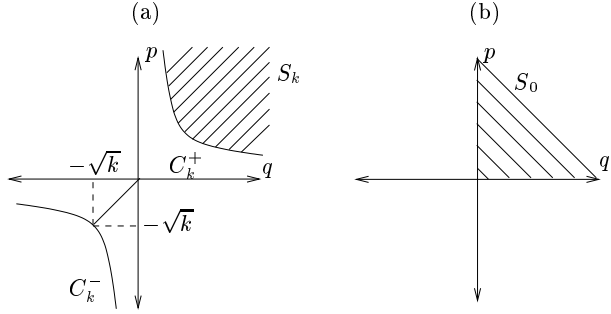,width=85mm}}
\caption{Graphs of hyperbolic regions and their boundaries: in (a), the
hyperbolic region $S_k$ is shown as are the boundary curves $C_k^+$ and
$C_k^-$, while in (b), the infinite wedge $S_0$ is depicted.}
\label{onecurve}
\end{figure}

Now consider the subregion $S_k$ that contains all points in $\Gamma$ such that
$qp \geq k,\, q \geq 0$, which is indicated by the shaded region in part (a) of
Figure \ref{onecurve}. Since this region can be viewed as the union of all
hyperbolic curves $C_l^+$ with $l\geq k$, it is itself invariant under the
action of $T_\sigma$. In order to apply this symmetry to the problem of
determining the bounds on qpis over $S_k$, we must first construct the
corresponding region operator, which we denote by $\hC_k$. Note that the
characteristic function on $S_k$, may be written as
\begin{equation}
\chi_k \,=\, \left\{ \begin{array}{cc} 1 & qp \geq k\,, q\geq 0 \\ 0  & \tO
\end{array} \right. \,=\, H(q)\,H(p-k/q) \,,
\label{bpb02}
\end{equation}
where $H$ is the Heaviside function. Using (\ref{bqi04b}), the configuration
kernel for the region operator can be determined (see the Appendix for
details):
\begin{equation}
\chi_{k,K}(x,y) \,=\, H(\tfrac{x+y}{2})e^{2ik \frac{x-y}{x+y}} \left[
\tfrac{1}{2} \delta(x-y) - \frac{1}{2\pi i (x-y)} \right]\,,
\label{qhp516}
\end{equation}
and hence the bounds on qpis are given by the spectral bounds associated with
the integral equation 
\begin{equation}
\int_{-x}^\infty e^{2ik \frac{x-y}{x+y}} \left[
\tfrac{1}{2} \delta(x-y) - \frac{1}{2\pi i (x-y)} \right] \psi(y) dy \,=\, \mu
\psi(x)\,.
\label{bpb03}
\end{equation}

We know, however, that the region operator $\hC_k$ is invariant under the set
of operator transformations that correspond to the subgroup of $Mp(2,\mathbb{R})$
formed by $T_\sigma, \sigma >0$. Since the effect of $T_\sigma$ is to squeeze
position and stretch momentum (or vice versa) while preserving the canonical
commutation relations, the corresponding operator transformation, up to an
unimportant phase, is given by the squeezing operator $\hat{U}_\sigma =
\exp(i\log \sigma (\hat{q} \hat{p} + \hat{p} \hat{q})/2)$. This implies that
$\hC_k$ commutes with $\hat{U}_\sigma$ for all $\sigma >0$, and hence
\begin{equation*}
[\hC_k,\hat{\omega}] \,=\, 0 \,, \quad \hat{\omega} \,=\, (\hat{q} \hat{p} +
\hat{p} \hat{q})/2\,. 
\end{equation*} 
It follows that the eigenfunctions of $\hC_k$ can be chosen such that they are
also eigenfunctions of $\hat{\omega}$. We can then obtain a partial solution to
the integral equation (\ref{bpb03}) by solving the equation $\hat{\omega} \psi
= \omega \psi$. A number of results connected with this problem can be found in
a paper of Chruscinski \cite{Chruscinski03}. On configuration space, this
equation appears as the first order differential equation
\begin{equation}
x\frac{d\psi}{dx} \,=\, (i\omega-\tfrac{1}{2})\psi\,.
\label{qhp506}
\end{equation}
The solutions of this equation are complex-valued linear combinations of the
functions
\begin{equation}
\psi_\omega^+(x) \,=\, \left\{ \begin{array}{cc} \frac{1}{\sqrt{2\pi}}
\frac{e^{i\omega\log|x|}}{|x|^{1/2}} & x>0  \\ 0 & x<0   \end{array}
\right. \,, \quad 
\psi_\omega^-(x) \,=\, \left\{ \begin{array}{cc} 0 & x> 0\,, \\
\frac{1}{\sqrt{2\pi}} \frac{e^{i\omega\log|x|}}{|x|^{1/2}} & x<0 \,.
\end{array} \right. \label{qhp507}
\end{equation}
Here $\omega$ can take any real value. These solutions are generalized
functions and are elements of the space of tempered distributions $\MG'$
\cite{Gelfand64}, of which  $\MH$ is a proper subspace. The factor
$1/\sqrt{2\pi}$ is inserted to ensure that
$(\psi_\omega^\pm,\psi_{\omega'}^\pm) = \delta(\omega - \omega')$. Since they
have disjoint support, $\psi_\omega^+$ and $\psi_{\omega'}^-$ are orthogonal
for all $\omega, \omega' \in \mathbb{R}$.  Note that, since $\log |x|
\rightarrow -\infty$ as $|x| \rightarrow 0$, the eigenfunctions become highly
oscillatory in the neighbourhood of the origin and are undefined at $|x|=0$,
due to the $|x|^{1/2}$ term in the denominator. 

Since the $\psi_\omega^\pm$ form two independent families of solutions to
(\ref{qhp506}), the eigenfunctions of $\hC_k$ are not yet fully determined. In
order to construct these solutions, we must solve the reduced eigenvalue
problem
\begin{equation}
\hC_k \psi_\omega\,=\, \mu(\omega,k) \psi_\omega\,, \quad \psi_\omega
\,=\, \alpha_\omega \psi_\omega^+ + \beta_\omega \psi_\omega^-\,, 
\label{qhp510}
\end{equation}
where $\alpha_\omega, \beta_\omega \in \mathbb{C}$. In order to solve
(\ref{qhp510}), we must first determine the action of $\hC_k$ on the
two-dimensional subspace $\MG'_\omega$ of $\MG'$ spanned by $\psi_\omega^+$ and
$\psi_\omega^-$: $\hC_k (\psi_\omega^+,\psi_\omega^-)^T =
A(\omega,k)(\psi_\omega^+,\psi_\omega^-)^T$, where $A(\omega,k)$ is given by
the matrix  
\begin{equation}
A(\omega,k) \,=\, \left( \begin{array}{cc} A_{11}(\omega,k) &
A_{12}(\omega,k) \\ A_{21}(\omega,k) & A_{22}(\omega,k)
\end{array}\right)\,.
\label{qhp511}
\end{equation}

The matrix elements of $A(\omega,k)$ can be computed by using the configuration
realization of $\hC_k$, details of which are presented in the Appendix. It so
happens that the matrix elements of $A(\omega,k)$ depend on the
functions $d(\omega,k)$ and $a(\omega,k)$, that are given by
\begin{equation}
d(\omega,k) \,=\, \frac{1}{2}\left[ \tanh(\pi \omega) + \frac{1}{4\pi}\Im\left\{
\oint_{C_0} \frac{e^{i\omega z - 2ik\coth(z/2)}}{\cosh(z/2)} dz\right\}\right]\,,
\label{qhp517d}
\end{equation}
where $C_0$ is any closed path in the complex plane that contains only the pole
at $z=0$, and 
\begin{eqnarray}
a(\omega,k) &=& \frac{e^{-\pi \omega}}{2\pi}\left( \int_0^\pi \left(
\frac{e^{\omega t - 2k\tan(t/2)}}{\sin(t/2)} - \frac{\cos(\omega t -
2k\tanh(t/2))}{\sinh(t/2)} \right)dt \right. \nonumber \\
& -& \left. \int_\pi^\infty \frac{\cos(\omega t - 2k\tanh(t/2))}{\sinh(t/2)} dt
\right)\,. \label{qhp517a}
\end{eqnarray}
The above formula is written in this way, because the individual terms in the
first integral are singular at $t=0$, whereas their difference is well-defined.
In terms of these functions, we may expand $A(\omega,k)$ as
\begin{equation}
\begin{array}{lcr} A(\omega,k) &=& \left( \begin{array}{cc} \tfrac{1}{2} +
d(\omega,k) & \tfrac12 \left[a(\omega,k) + ie^{-\pi\omega}(\tfrac12 +
d(\omega,k)) \right] \\ \tfrac12 \left[ a(\omega,k) - ie^{-\pi\omega}(\tfrac12 +
d(\omega,k)) \right] & 0 \end{array} \right) \,. \end{array}
\label{qhp540}
\end{equation}
Hence the spectrum of the region operator $\hC_k$ splits into positive and
negative parts, which we label by $\mu_+(\omega,k)$ and $\mu_-(\omega,k)$
respectively:
\begin{equation}
\mu_\pm(\omega,k) \,=\, \frac12\left[ \frac12 + d(\omega,k) \pm \sqrt{
(\tfrac12 + d(\omega,k))^2(1+e^{-2\pi\omega}) + a(\omega,k)^2} \right]\,,
\label{qba502}
\end{equation}
Of particular interest are the functions $L_k=\inf_{\omega\in
\mathbb{R}}\mu_-(\omega,k)$ and $U_k=\sup_{\omega \in \mathbb{R}}
\mu_+(\omega,k)$, since they provide the best-possible bounds on qpis over the
hyperbolic regions $S_k$. Although it does not seem possible to obtain exact
expressions for these functions, it is not difficult to compute the bounds
after first evaluating $a$ and $d$ numerically. 

These bounds are graphed in Figure \ref{fig5h06} for $k$ in the range $[0,5]$
from which we conclude that the upper bound on qpis over $S_k$ remains close to
but greater to $1$ for all $k$ and that this difference is greatest when $k=0$
(see inset). The lower bound displays a more marked difference from the
classical bound of $0$, reaching a minimum value of $-0.3089$ at approx.
$k=1.9$, before rising again. A surprising result is that the lower bound does
not approach $0$ for large values of $k$. Nonetheless, this appears to be a
characteristic feature of bounds on qpis for many classes of regions
\cite{Wood04}. 

\begin{figure}
\centerline{\psfig{figure=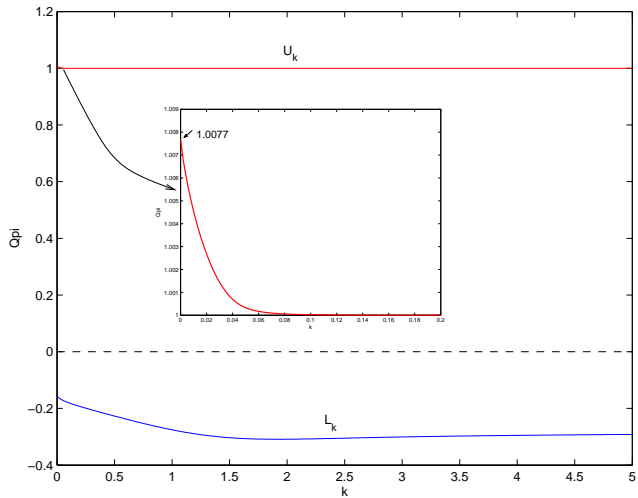,width=85mm}}
\caption{Graph of the best-possible bounds on qpis over $S_k$ for $k$ in the
range $0$ to $5$. The inset graph, with $k$ in the range $0$ to $0.2$, shows
that the upper bound lies above $1$, but converges rapidly to $1$ as $k$
increases.}
\label{fig5h06}
\end{figure}

\begin{subsection}{THE INFINITE WEDGE}

An interesting subclass of hyperbolic regions is provided by taking the limit
as $k\rightarrow 0$. The region $S_0$ so obtained  is precisely the positive
$(q,p)$ quadrant of $\Gamma$, as depicted in part (b) of Figure \ref{onecurve}.
Note that when $k=0$, the functions $d$ and $a$ take a simplified form:
\begin{equation}
d(\omega,0)) \,=\, \tfrac12 \tanh(\pi \omega)\,, \quad a(\omega,0) \,=\,
-\tfrac12 \tanh(\pi \omega) + u(\omega)\,,
\label{bpb05}
\end{equation}
where $u(\omega)$ may be expressed as an infinite sum \cite{Gradshteyn00}:
\begin{equation*}
u(\omega) \,=\, \frac{8\omega}{\pi} \sum_{n=0}^\infty
\frac{1}{\omega^2+(4n+1)^2}\,. 
\end{equation*}
If we now apply these simplifications to the spectral formula (\ref{qba502}),
we obtain the spectrum for $\hC_0$:
\begin{equation}
\mu_\pm(\omega,0) \,=\, \tfrac14 \left(1 +\tanh(\pi \omega) \pm
\sqrt{(2u(\omega) - \tanh(\pi \omega))^2 + (1+\tanh(\pi \omega))^2}\right)\,,
\label{qbw509}
\end{equation}
which is graphed in Figure \ref{fig5h09}. The infimum and supremum can then be
determined numerically, and to an accuracy of $\pm 5\times 10^{-10}$, we have
that
\begin{equation}
-0.155939843 \,<\, Q_0[W] \,<\, 1.007679970.
\label{bpb06}
\end{equation}

\begin{figure}
\centerline{\psfig{figure=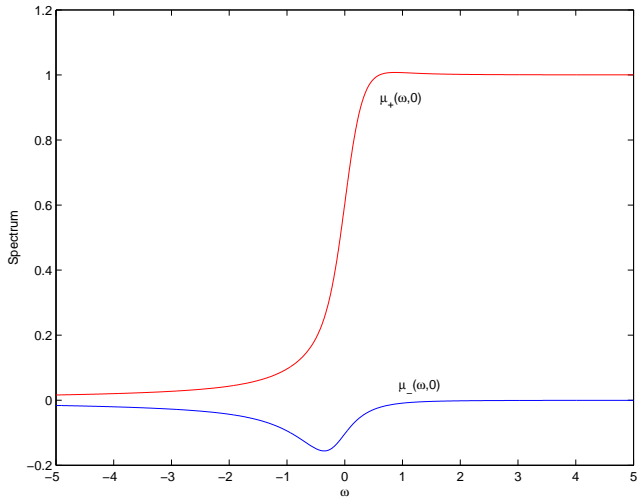,width=85mm}}
\caption{Graph of the spectrum of $\hC_0$ which, as labeled, splits into the
curves $\mu_+(\omega,0)$ and $\mu_-(\omega,0)$}
\label{fig5h09}
\end{figure}

An interesting point is that these bounds are also best-possible when applied
to regions defined by infinite wedges. This equivalence is due to two factors:
firstly, by an appropriate metaplectic transformation $T$, the region $S_0$ can
be transformed into any infinite wedge with half-angle $\alpha < \pi/2$.
Secondly, the operator transformation $\hat{U}_T$ that corresponds to $T$ is
unitary, and thus the spectrum of $\hC_0$ is preserved under its action. This
implies that the spectrum of any region operator corresponding to an infinite
wedge is given by (\ref{qbw509}). As a consequence, the integral of a Wigner
function over any infinite wedge must lie between the bounds given in
(\ref{bpb06}). Bounds on the spectrum of similar operators have been considered
before, in the context of the quantum phase operator \cite{Dubin00} and in
connection with studies of probability backflow \cite{Bracken94} but not to the
same level of precision. 

\end{subsection}
\end{section}

\begin{section}{EXAMPLES INVOLVING TWO BOUNDARY CURVES}

As a second example, consider the slightly more complicated case of a region
with a boundary composed of two curves with $T_\sigma$ symmetry. There are
several possible forms that such a region can take \cite{Wood04}, however, we
will concentrate on just the subcase for which the boundary curves lie in
positive and negative $(q,p)$ quadrants, as shown in part (a) of Figure
\ref{twocurve}. 

\begin{figure}
\centerline{\psfig{figure=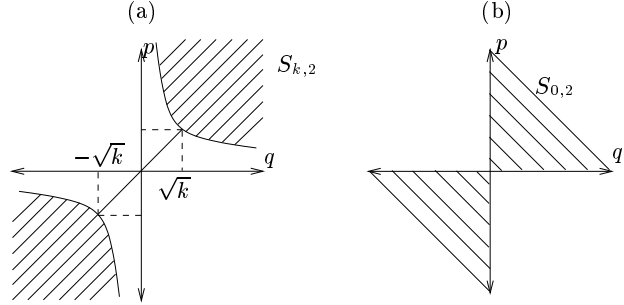,width=85mm}}
\caption{Regions bounded by two hyperbolic curves: in (a), the hyperbolic
region $S_{k,2}$ is shown while in (b) the double wedge $S_{0,2}$ is depicted.}
\label{twocurve}
\end{figure}

In order to further simplify matters, we assume that both curves are labeled
by the variable $k$. We label the class of regions that remain by $S_{k,2}$ and
note that this region may be written in terms of the region $S_k$ as
\begin{equation}
S_{k,2} \,=\, S_k + R_\pi (S_k)\,,
\label{bpc01}
\end{equation}
where $R_\pi$ denotes a rotation through an angle $\pi$. Note that $R_\pi
:(q,p) \rightarrow (-q,-p)$ and that the operator that corresponds to this
transformation under the Weyl-Wigner transform is just the parity operator
$\hat{P}$, which acts on the canonical coordinate and momentum operators
according to
\begin{equation*}
\hat{P}\hat{q} \hat{P} \,=\, -\hat{q}\,, \quad \hat{P} \hat{p} \hat{P} \,=\,
-\hat{p}\,.
\end{equation*}
Due to the linearity of the Weyl quantization map, this implies that the
region operator that corresponds to $S_{k,2}$ may be expressed as
\begin{equation}
\hC_{k,2} \,=\, \hC_k + \hat{P} \hC_k \hat{P}\,.
\label{bpc03}
\end{equation}
This operator also commutes with $\hat{\omega}$, and hence its eigenstates can
be chosen such that they are some linear combination of $\psi_\omega^\pm$. In
order to find the correct linear combination, we must first determine the
matrix representation of $\hC_{k,2}$ on the subspace spanned by
$\psi^\pm_\omega$. This turns out to quite simple, since the action of
$\hat{P}$ on this subspace is given by $\hat{P} \psi_\omega^\pm =
\psi_\omega^\mp$. Thus the matrix representation of $\hC_{k,2}$ is given by
\begin{equation}
A_2(\omega,k) \,=\, A(\omega,k) + \left( \begin{array}{cc} 0 & 1 \\ 1 & 0
\end{array} \right) A(\omega,k) \left( \begin{array}{cc} 0 & 1 \\ 1 & 0
\end{array} \right) \,=\, \left( \begin{array}{cc} \tfrac12 + d(\omega,k) &
a(\omega,k) \\ a(\omega,k) & \tfrac12 + d(\omega,k) \end{array} \right)\,.
\label{bpc04}
\end{equation}

The simple form of this matrix representation leads to the following expression
for the spectrum of $\hC_{k,2}$:
\begin{equation}
\mu_{2,\pm}(\omega,k) \,=\, \frac12 + d(\omega,k) \pm |a(\omega,k)|\,,
\label{bpc05}
\end{equation}
In this case, the eigenfunctions are odd and even combinations of
$\psi_\omega^\pm$, and are independent of $k$: 
\begin{equation}
\psi^{\pm}_{2,\omega}(x;k) \,\equiv \psi^{\pm}_{2,\omega}(x) \,=\,
\frac1{\sqrt{2}}\left(\psi_\omega^+(x) \pm \psi_\omega^-(x) \right) \,, \label{bpc06}
\end{equation}
which indicates that the operators $\hC_{k,2}$ commute for all $k\geq 0$.

The properties of the spectrum in this case vary somewhat from the preceding
example. In particular, $\mu_{2,-}(\omega,k)$ is not restricted to negative
values and, similarly, $\mu_{2,+}(\omega,k)$ is not strictly positive, although
clearly the inequality $\mu_{2,+}(\omega,k) \geq \mu_{2,-}(\omega,k)$ holds for
all $\omega \in \mathbb{R}, k \geq 0$. Since the bounds on qpis over $S_{k,2}$
are given by the infimum $L_{k,2}$ and supremum $U_{k,2}$ of the spectrum of
$\hC_{k,2}$, it is these functions that are of primary importance in the
context of this paper. Again, closed-form expressions do not appear to exist,
so we must resort to computational techniques in evaluating these functions,
the results of which are graphed in Figure \ref{fig5h17}. In this case, the
upper bound is well in excess of $1$ for small values of $k$, but rapidly
approaches $1$ as $k$ increases. The lower bound, on the other hand, dips
initially, reaching a minimum of $-0.4014$ at $k=0.4$ before rising again, and
appears to approach a finite negative value near to $-0.3$ as $k \rightarrow
\infty$. 

\begin{figure}
\centerline{\psfig{figure=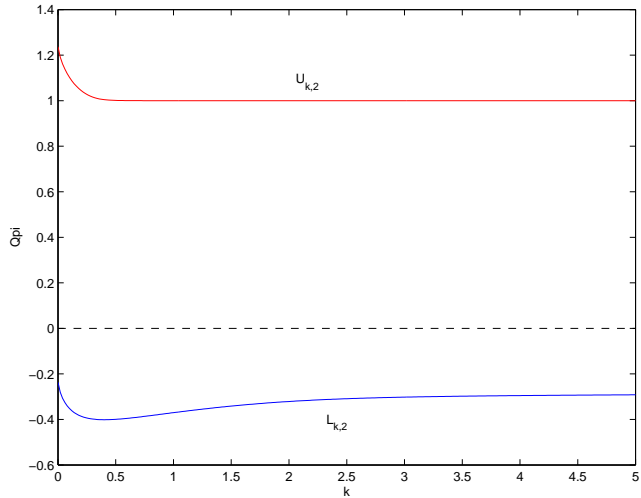,width=85mm}}
\caption{Graph of the best-possible bounds on qpis over $S_{k,2}$ for $k$ in
the range $0$ to $5$.}
\label{fig5h17}
\end{figure}

\begin{subsection}{DOUBLE WEDGES}

It is again of interest to consider in more detail the limit as $k\rightarrow
0$ of the region $S_{k,2}$. The resulting region $S_{0,2}$ is the union of the
positive and negative $(q,p)$ quadrants (as shown in part (b) of Figure
\ref{twocurve}, and one might guess that qpis over such a region should be
positive \cite{Bertrand87}, since it appears to composed from the union of a
set of infinite straight lines, over which the integral of the Wigner function
is known to be positive. However, since these lines cross at the origin one
cannot immediately apply this result and it will be shown that the true bounds
on qpis lie significantly outside the $[0,1]$ interval to which classical
probabilities are restricted.

The region operator that corresponds to $\hC_{0,2}$ may be expressed in terms
of $\hC_0$ as $\hC_{0,2} = \hC_0 + \hat{P}\hC_0\hat{P}$. The spectrum for this
operator can be derived from (\ref{bpc05}) upon substitution of (\ref{bpb05}),
from which we obtain
\begin{equation}
\mu_{2,\pm}(\omega,0) \,=\, \frac{1+\tanh(\pi\omega)}{2} \pm \left| u(\omega) -
\frac{\tanh(\pi\omega)}{2} \right|\,.
\label{bpc08}
\end{equation}
This spectrum is graphed in Figure \ref{fig5h19}, and from this one sees that
the function $\mu_2 = \tfrac12 + u(\omega)$ passes through both the infimum and
the supremum of the spectrum of $\hC_{0,2}$. Since $u(\omega)$ is an odd
function, we need only calculate its global maximum in order to determine the
bounds on qpis over $S_{0,2}$. Using computational techniques, this value can
be obtained to great accuracy, and we find that the best-possible bounds
(accurate to $\pm 5 \times 10^{-10}$) on qpis over $S_{0,2}$ are
\begin{equation}
-0.236823652 \,<\, Q_{\alpha,2}[W] \,<\, 1.236823652\,.
\label{qbe508}
\end{equation}
Note that the upper and lower bounds sum to $1$ since they are symmetric about
$1/2$. This symmetry can be explained by noting that if one rotates the region
$S_{0,2}$ through an angle $\pi/2$, then one obtains its complement: i.e.
$R_{\pi/2} (S_{0,2}) = \Gamma / S_{0,2} = S_{0,2}^c$. Note that the integral of
a Wigner function over $\Gamma=S\cup S^c$ is normalized to $1$. Now, since the
operator equivalent of a rotation is a unitary transformation, the region
operator that corresponds to the complement of $S_{0,2}$ has precisely the
spectrum given in (\ref{bpc08}). Accordingly, the spectrum of $\hC_{0,2}$ must
consist of pairs that sum to $1$ and, in particular, the upper and lower
bounds on this spectrum must be symmetric about $1/2$.

\begin{figure}
\centerline{\psfig{figure=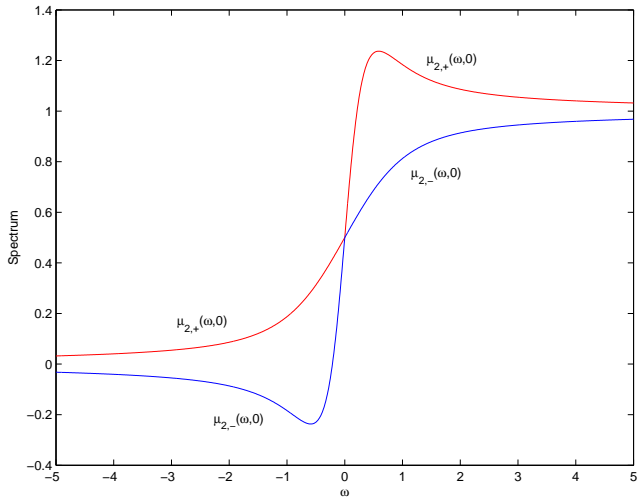,width=85mm}}
\caption{Spectrum of the double wedge operator $\hC_{0,2}$.}
\label{fig5h19}
\end{figure}

As in the case of the region $S_0$, the bounds on qpis over $S_{0,2}$ can be
applied to a wider class of regions. We shall refer to elements of this wider
class as double wedges, since they are formed by taking the union of an
infinite wedge with its rotation through an angle $\pi$. By applying the
appropriate metaplectic transformation, we can transform $S_{0,2}$ into any
double wedge. The corresponding operator transformation is unitary and preserves
the spectrum of $\hC_{0,2}$, so that the spectrum of any region operator
corresponding to a double wedge is given by (\ref{bpc08}). Accordingly, the
integral of any Wigner function over an arbitrary double wedge must satisfy the
inequality given in (\ref{qbe508}).

\end{subsection}
\end{section}

\begin{section}{CONCLUSION}
\end{section}

The problem of constructing best possible bounds on integrals of the Wigner
function is not only of mathematical interest, but should be of practical
significance in providing checks on experimentally reconstructed quantum
states. Since our approach to the problem relies on specifying the region to be
integrated over, it is important to identify the types of region for which the
bounds can be easily computed. In this paper, we have considered several
examples of regions with a hyperbolic symmetry for which the bounds can be
computed numerically from the spectrum of an exactly solvable integral
equation. We have demonstrated that the bounds on integrals of the Wigner
function for these regions are not equivalent to those on integrals of true
probability density functions. In particular, the lower bound is significantly
below zero in all cases, although it lacks the scalloped effect arising from
eigenvalue crossings as seen in the bounds for elliptical discs
\cite{Bracken99}. The upper bound also rises above $1$ although for the most
part the difference between its value and the classical bound is very small.
This contrasts with the case of the disc, for which the upper bound always
remains below $1$.

\begin{figure}
\centerline{\psfig{figure=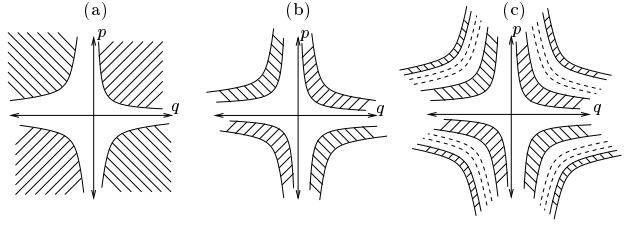,width=85mm}}
\caption{Examples of generalized hyperbolic regions are shown in (a)-(c).}
\label{gencurve}
\end{figure}

The results herein can also be extended to more complicated regions with
boundaries given by an arbitrary number of hyperbolic curves sharing the same
symmetry \cite{Wood04}, for example, the regions shown in Figure
\ref{gencurve}. The problem of determining the spectrum is essential the same
but the matrix representations for operators corresponding to regions with many
boundaries are functions of many variables and hence the behaviour of the
bounds is much more difficult to characterize.

\newpage

\renewcommand{\theequation}{A-\arabic{equation}}
\setcounter{equation}{0}  
\section*{APPENDIX}  

The configuration kernels that correspond to (\ref{bpb02}) take the form
\begin{equation}
\chi_{k, K}(x,y) \,=\, \frac{H(\tfrac{x+y}{2})}{2\pi}
\int_{-\infty}^\infty H(p-\tfrac{2k}{x+y}) e^{i(x-y)p} dp\,.
\label{ahoa02}
\end{equation}
This integral can be computed in a generalized sense \cite{Sneddon51}, and we
find that
\begin{equation}
\chi_{k,K}(x,y) \,=\, H(\tfrac{x+y}{2})e^{2ik \frac{x-y}{x+y}} \left[ \tfrac{1}{2}
\delta(x-y) - \frac{1}{2\pi i (x-y)} \right]\,.
\label{ahoa05}
\end{equation}

This expression for the configuration kernel of $\hC_k$ enables us to determine
the action of $\hC_k$ on the space $\MG'_\omega$ (recall that this is given by
the matrix $A(\omega,k)$ defined in (\ref{qhp511})) . In this representation
$\hC_k$ acts on $\psi_\omega$ as
\begin{equation}
(\hC_k \psi_\omega)(x) \,=\, \int_{-x}^\infty
e^{2ik\frac{x-y}{x+y}} \left[\tfrac12 \delta(x-y) - \frac{1}{2\pi i
(x-y)}\right] \psi_\omega(y) dy\,.
\label{ahoa08}
\end{equation}
If we substitute $\psi_\omega=\alpha_\omega\psi_\omega^+ +
\beta_\omega\psi_\omega^-$, then we find that the action of $\hC_k$
when $x<0$ differs from that when $x>0$. Thus, for $x>0$,
\begin{eqnarray*}
(\hC_k \psi_\omega)(x) &=& \alpha_\omega \int_0^\infty
e^{2ik\frac{x-y}{x+y}} \left[\tfrac12 \delta(x-y) - \frac{1}{2\pi i
(x-y)}\right] \frac{e^{i\omega\log y}}{\sqrt{2\pi y}} dy  
\\ &+& \beta_\omega \int_{-x}^0
e^{2ik\frac{x-y}{x+y}} \left[\tfrac12 \delta(x-y) - \frac{1}{2\pi i
(x-y)}\right] \frac{e^{i\omega\log |y|}}{\sqrt{2\pi |y|}} dy  
\end{eqnarray*}
and for $x<0$,
\begin{equation*}
(\hC_k \psi_\omega)(x) \,=\, \alpha_\omega \int_{|x|}^\infty
e^{2ik\frac{x-y}{x+y}} \left[\tfrac12 \delta(x-y) - \frac{1}{2\pi i
(x-y)}\right] \frac{e^{i\omega\log y}}{\sqrt{2\pi y}} dy \,.
\end{equation*}
It is immediately clear that $A_{22}(\omega,k)=0$, since the $x>0$ case
involves only $\alpha_\omega$.

The other integrals can be simplified and this process leads to the following
expression for the matrix elements of $A$:
\begin{equation}
A(\omega,k)\, =\, \left( \begin{array}{cc} \tfrac{1}{2} +
d(\omega,k) & \tfrac{1}{2}[a(\omega,k) + ib(\omega,k)] \\
\tfrac{1}{2}[a(\omega,k) - ib(\omega,k)] & 0 \end{array} \right) \,,
\end{equation}
where the functions $d, a$ and $b$ are given by
\begin{eqnarray}
d(\omega,k) &=& \frac{1}{2\pi} \int_0^\infty \frac{\sin(\omega t -
2k\tanh(t/2))}{\sinh(t/2)} dt\,, \label{qhp517}\\
a(\omega,k) &=& \frac{1}{2\pi} \int_0^\infty \frac{\sin(\omega t -
2k \coth(t/2))}{\cosh(t/2)} dt\,,  \label{qhp518} \\
b(\omega,k) &=& \frac{1}{2\pi} \int_0^\infty \frac{\cos(\omega t -2k
\coth(t/2))}{\cosh(t/2)} dt.
\label{qhp519} 
\end{eqnarray}
Note, however, that although the integrands of $a$ and $b$ are bounded for all
$t$, they become highly oscillatory in the neighbourhood of the origin, which
poses difficulties for numerical schemes. These problems can be alleviated by
using the technique of contour integration.

\begin{figure}
\centerline{\psfig{figure=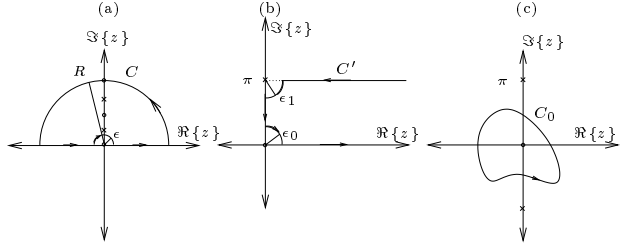,width=85mm}}
\caption{The contours used in evaluating the functions
$b(\omega,k),a(\omega,k)$ and $d(\omega,k)$: in (a), the semi-circular contour
$C$, in (b), the contour $C'$ used to relate $b(\omega,k)$ to $d(\omega,k)$ and
in (c), an example of a contour $C_0$ for evaluating $R(\omega,k)$.}
\label{contour1}
\end{figure}

In the case of $b(\omega,k)$, we consider the following contour integral in the
complex plane
\begin{equation}
I_{C} = \frac{1}{2\pi} \oint_C \frac{e^{i\omega z -
2ik\coth(z/2)}}{\cosh(z/2)} dz\,,
\label{cont01}
\end{equation}
where $C$ is the contour shown in part (a) of Figure \ref{contour1}. Although
the contour is divided into four parts, only the integrals along the real axis
contribute, since the contributions from the semi-circular segments vanish in
the respective limits as $\epsilon \rightarrow 0$ and $R \rightarrow \infty$.
Thus one has that
\begin{equation}
\lim_{R\rightarrow\infty,\epsilon \rightarrow 0} I_C \,=\, 2 \frac{1}{2\pi}
\int_0^{\infty} \frac{e^{i\omega t - 2ik\coth(t/2)}}{\cosh(t/2)}
dt\,,
\label{cont02}
\end{equation}
and as a result, $b(\omega,k) \,=\, \Re\{ I_C\}/2$.

We can make use of the residue theorem in evaluating $I_C$:
\begin{equation}
I_C \,=\, 2\pi i\sum \textrm{Res} f(z)\,,
\label{cont04}
\end{equation}
where $f(z)$ is the integrand in (\ref{cont01}). Note that $f(z)$ has two
distinct classes of residues: simple poles at $z=(2n+1)\pi i$ and essential
singularities at $z=2m\pi i$, with $m \in \mathbb{Z}$. The contour $C$ encloses
only the simple poles with $n\geq 0$ and the essential singularities with $m
\geq 1$, and hence the sum in (\ref{cont04}) is over the residues at these
points.

It is easy to evaluate the residues at the simple poles and we find that the
total contribution from the simple poles inside $C$ is given by
\begin{equation}
\textrm{Res}_s \,=\, \frac{1}{2\pi i}\sum_{n=0}^\infty (-1)^n e^{-(2n+1)\pi
\omega} \,=\, \frac{\sech(\pi \omega)}{2\pi i}\,.
\label{cont05}
\end{equation}
The sum of the residues associated with the essential singularities can also be
simplified:
\begin{equation}
\textrm{Res}_e \,=\, \frac{1}{2\pi} \sum_{n=1}^\infty (-1)^n e^{-2n\pi \omega}
R(\omega,k)\,=\, =\frac{e^{-\pi\omega}\sech(\pi \omega)}{2} R(\omega,k)\,,
\label{cont06}
\end{equation}
where $R(\omega,k)$ is the residue of $f(z)$ associated with the essential
singularity at the origin.

Ordinarily, one might try to evaluate this residue by constructing the Laurent
series for $f(z)$, however in this case the $\coth$ term in the exponential
makes this extremely difficult. However, one may use the residue theorem in
reverse and evaluate $R(\omega,k)$ by considering the integral of $f(z)$ over a
closed contour enclosing the origin (and no other poles):
\begin{equation}
R(\omega,k) \,=\, \frac{1}{2\pi i} \oint_{C_1} \frac{e^{i\omega z -
2ik\coth(z/2)}}{\cosh(z/2)} dz\,.
\label{cont07}
\end{equation}
The rapid oscillations due to the $\coth$ term do not appear in this
calculation, and due to its finite range this integral can be rapidly evaluated
to a high degree of accuracy using numerical techniques.

If we now collect the results for the residues together, we discover that
\begin{equation}
b(\omega,k) \,=\, \frac{\sech(\pi \omega)}{2} \left[ 1 + \frac{e^{-\pi \omega}}{2}
\Im\{R(\omega,k)\} \right]\,.
\label{cont08}
\end{equation}
Note that in deriving these results it has been assumed that $\omega \geq 0$. A
similar procedure (with the semi-circular contour defined in the lower half
plane) enables us to extend the validity of (\ref{cont08}) to all $\omega \in
\mathbb{R}$.

In order to obtain superior expressions for the functions $a(\omega,k)$ and
$d(\omega,k)$, we choose another contour (see part (b) of Figure
\ref{contour1}), this time involving five curves. However, we know from the
above calculation that the integral over $\epsilon_0$ vanishes, which leaves
four curves to consider. Of these, the integral over the imaginary axis from
$0$ to $\pi$ results in a pure imaginary contribution $I_\pi$, while the
integral over $\epsilon_1$ contributes $-1/2$ in the limit as $\epsilon_1
\rightarrow 0$. The integral over the positive real axis is equal to
$b(\omega,k)+a(\omega,k)i$ in the limit as $\epsilon_0 \rightarrow 0$, while
the contribution from the line $(\pi i,\pi i + \infty)$ can be expressed as
$-\exp(-\pi \omega)(d(\omega,k) + I_\infty i)$. 

By equating the real parts, we find that
\begin{equation}
b(\omega,k) \,=\, e^{-\pi \omega} (\tfrac{1}{2} + d(\omega,k))\,.
\end{equation}
This leads to the expression (\ref{qhp517d}) for $d(\omega,k)$ and the
expression (\ref{qhp540}) for the matrix $A$. If we equate the imaginary parts,
then we discover that
\begin{equation}
a(\omega,k) \,=\, e^{-\pi \omega} (I_\pi - I_\infty)\,,
\end{equation}
where
\begin{eqnarray}
I_\pi &=& \frac{1}{2\pi}\int_0^\pi \frac{e^{\omega t-2k tan(t/2)}}{\sin(t/2)}
dt\,, \\
I_\infty &=& \frac{1}{2\pi}\int_0^\infty \frac{\cos(\omega t -
2k\tanh(t/2))}{\sinh(t/2)} dt\,.
\end{eqnarray}
Neither $I_\pi$ nor $I_\infty$ are well-defined, but their difference is, and
provided one expresses $a(\omega,k)$ as in (\ref{qhp517a}), the singularities of
these integrals are avoided.

\newpage

\renewcommand{\bibnumfmt}[1]{$^{#1}$}

\end{document}